%% file: aaai25.tex
\documentclass[letterpaper]{article} 
\usepackage{aaai25}  
\usepackage{times}  
\usepackage{helvet}  
\usepackage{courier}  
\usepackage[hyphens]{url}  
\usepackage{graphicx} 
\usepackage{natbib}  
\usepackage{caption} 
\usepackage{subcaption}

\newtheorem{definition}{Definition} 
\usepackage{amsmath}
\usepackage{multirow}
\usepackage{bm}
\usepackage{booktabs}
\usepackage{amsfonts}
\usepackage{amssymb}

\usepackage{algorithm}
\usepackage{algorithmic}
\usepackage{xcolor}
 
\usepackage{newfloat}
\usepackage{listings}
\DeclareCaptionStyle{ruled}{labelfont=normalfont,labelsep=colon,strut=off} 
\floatstyle{ruled}
\newfloat{listing}{tb}{lst}{}
\floatname{listing}{Listing}
\nocopyright

\title{Harnessing Multimodal Large Language Models for Multimodal Sequential Recommendation}
\author{
Yuyang Ye\textsuperscript{\rm 1},Zhi Zheng\textsuperscript{\rm 2}, Yishan Shen\textsuperscript{\rm 3}, Tianshu Wang\textsuperscript{\rm 4}, Hengruo Zhang\textsuperscript{\rm 4}, \\Peijun Zhu\textsuperscript{\rm 5}, Runlong Yu\textsuperscript{\rm 6}, Kai Zhang\textsuperscript{\rm 2}, Hui Xiong\textsuperscript{\rm 7}\thanks{Corresponding Author}
}
\affiliations{
    \textsuperscript{\rm 1}Rutgers University,
    \textsuperscript{\rm 2}University of Science and Technology of China,
    \textsuperscript{\rm 3}University of Pennsylvania\\
    \textsuperscript{\rm 4}Bytedance Inc.,
    \textsuperscript{\rm 5}Georgia Institute of Technology,
    \textsuperscript{\rm 6}University of Pittsburgh,\\
    \textsuperscript{\rm 7}The Hong Kong University of Science and
Technology.\\
    yuyang.ye@rutgers.edu, zhengzhi97@mail.ustc.edu, yishans@sas.upenn.edu, \{ustcwtsphy, hengruoz\}@gmail.com, zpj@gatech.edu, ryu@pitt.edu, kkzhang08@ustc.edu.cn, xionghui@ust.hk
}

\usepackage{bibentry}

\begin{document}

\maketitle

\begin{abstract}
Recent advances in Large Language Models (LLMs) have demonstrated significant potential in the field of Recommendation Systems (RSs). Most existing studies have focused on converting user behavior logs into textual prompts and leveraging techniques such as prompt tuning to enable LLMs for recommendation tasks. Meanwhile, research interest has recently grown in multimodal recommendation systems that integrate data from images, text, and other sources using modality fusion techniques. This introduces new challenges to the existing LLM-based recommendation paradigm which relies solely on text modality information. Moreover, although Multimodal Large Language Models (MLLMs) capable of processing multi-modal inputs have emerged, how to equip MLLMs with multi-modal recommendation capabilities remains largely unexplored. To this end, in this paper, we propose the Multimodal Large Language Model-enhanced Multimodal Sequential Recommendation (MLLM-MSR) model. To capture the dynamic user preference, we design a two-stage user preference summarization method. Specifically, we first utilize an MLLM-based item-summarizer to extract image feature given an item and convert the image into text. Then, we employ a recurrent user preference summarization generation paradigm to capture the dynamic changes in user preferences based on an LLM-based user-summarizer. Finally, to enable the MLLM for multi-modal recommendation task, we propose to fine-tune a MLLM-based recommender using Supervised Fine-Tuning (SFT) techniques. Extensive evaluations across various datasets validate the effectiveness of MLLM-MSR, showcasing its superior ability to capture and adapt to the evolving dynamics of user preferences. The code of our work is publicly available at https://github.com/YuyangYe/MLLM-MSR.
\end{abstract}

\input{sections/introduction}
\input{sections/related-work}

\input{sections/preliminary}

\input{sections/framework}

\input{sections/experiments}
\input{sections/conclusion}

\small{\bibliography{aaai25}}

\end{document}

%% file: sections/introduction.tex
\section{Introduction}
The development of Large Language Models (LLMs) has significantly enhanced the capacity for natural language understanding~\cite{floridi2020gpt, achiam2023gpt, touvron2023llama}, which has been instrumental in advancing recommendation systems (RSs). LLMs have demonstrated remarkable improvements in processing complex user preferences due to its strong semantic understanding and summarization ability. These attributes significantly enhance personalization and accuracy in recommendations\cite{wu2023survey, zhang2024notellm, ren2024representation}, particularly in Sequential Recommendations (SRs) where extracting long historical preferences is crucial~\cite{hou2023learning, zheng2024harnessing, zhai2023knowledge, li2023e4srec}.

Simultaneously, beyond solely modeling textual information, there has been a growing interest in leveraging multimodal information~\cite{liu2023multimodal}. Techniques such as multimodal fusion and gated multimodal units have been utilized to integrate data from various sources—images, videos, and audio—enriching the context for recommendations. This offers a deeper understanding of user-item interactions and naturally leads to the exploration of Multimodal Large Language Models (MLLMs) for enhancing multimodal recommendation systems~\cite{liu2021pre, zhou2024disentangled, kim2024monet}. MLLMs merge multimodal information into a unified textual semantic space, enhancing the system's ability to understand and interpret complex data inputs, thereby can significantly improving recommendation accuracy~\cite{liu2024rec, zhang2024notellm}. The application of MLLMs in sequential recommender systems presents a promising avenue for dynamically adapting to user preferences and handling the intricate interplay of multimodal data, which holds considerable untapped potential.

\par However, integrating Multimodal Large Language Models (MLLMs) into multimodal sequential recommendation systems introduces a set of notable challenges. First, the inherent complexity and computational demands of processing sequential multimodal data, particularly with multiple ordered image inputs, significantly constrain the scalability and efficiency of these systems \cite{yue2024mmmu,koh2024generating}. Moreover, conventional MLLMs often exhibit limitations in comprehending the temporal dynamics of user interactions and preferences, particularly in the context of sequential multimodal interactions \cite{gao2021clip, liu2024rec}. This critical limitation undermines the systems' capacity to accurately capture and reflect the evolving nature of user interests over time. Furthermore, fine-tuning multimodal large language models (MLLMs) for specific recommendation scenarios while avoiding overfitting and preserving the generalizability gained during pre-training presents a significant challenge \cite{borisov2022language, yin2023heterogeneous, li2023prompt}. These hurdles underscore the need for innovative approaches that can navigate the complexities of multimodal sequential data, ensuring that MLLMs can be effectively leveraged to enhance recommendation systems.

To address these challenges, this paper introduces the Multimodal Large Language Model-enhanced Multimodal Sequential Recommendation (MLLM-MSR), a pioneering approach that leverages the capabilities of MLLMs to enhance and integrate multimodal item data effectively. Specifically, we introduce a Multimodal User Preferences Inference approach, which merges traditional multimodal fusion with sequence modeling techniques with MLLMs. Initially, we employ MLLMs to transform visual and textual data of each item into a cohesive textual description, preserving the integrity of the information as demonstrated by a preliminary study. Subsequently, utilizing the enriched item information processed through the MLLM, we develop an innovative LLM-based recurrent method to infer user preferences, capturing the temporal dynamics of these preferences. This method addresses the above mentioned challenges in processing sequential image inputs by harnessing superior text process capabilities of LLMs and improves the interpretability of recommendation compared with traditional representation based approaches, by providing detailed user preference. Further, we fine-tune an MLLM to function as a recommender, utilizing a carefully designed set of prompts that integrate this enriched item data, inferred user preferences, and the ground-truth of user-item interactions. This process of Supervised Fine-Tuning (SFT) on an open-source MLLM, equips the model with the ability to accurately match user preferences with potential items, thereby enhancing the personalization and accuracy of recommendations. To validate the effectiveness of MLLM-MSR, we conduct extensive experiments across three publicly available datasets from various domains, which confirm the superior performance of our approach. The major contributions of this paper are summarized as follows:

\begin{itemize}
    \item To best of our knowledge, our work is the first attempt to fine-tune multimodal large models to address the challenges of sequential multimodal recommendation, where our fine-tune strategies achieving significant improvements in recommendation performance.
    \item We introduce a novel image summarizing method based on MLLMs to recurrently summarize user preferences on multi modality, facilitating a deeper understanding of user interactions and interests over time.
    \item Our approach is extensively validated across various datasets, demonstrating its effectiveness in enhancing the accuracy and interpretability of recommendations.
\end{itemize}

%% file: sections/related-work.tex
\section{Related Work}

\subsection{Multimodal Sequential Recommendation}
Sequential Recommenders (SRs) have progressed from matrix-based models to sophisticated neural architectures. Initially, Factorizing Personalized Markov Chains (FPMC) integrated matrix factorization with Markov chains to model sequential behavior \cite{rendle2010factorizing,yu2023cognitive}. The transition to neural models started with GRU4Rec, which employed gated recurrent units for session-based recommendations \cite{hidasi2015session}. Subsequently, SASRec utilized self-attention mechanisms to handle long-term dependencies in user-item interactions \cite{kang2018self}, and BERT4Rec introduced transformers to SRs, significantly enhancing performance with deep bidirectional training~\cite{sun2019bert4rec}.

Furthermore, the evolution of multimodal information-enhanced SRs has leveraged additional contextual information to improve recommendation quality. Fusion methods in SRs are categorized into early, late, and hybrid approaches~\cite{hu2023adaptive}. Early fusion techniques involve invasive methods that integrate various modalities at the input level, enhancing initial feature representation through techniques like concatenation and gating \cite{tang2018personalized, sun2019bert4rec, lei2019tissa}. Besides, non-invasive early fusion employs attention mechanisms to merge multiple attributes before processing \cite{rendle2019difsr, liu2021noninvasive}. In contrast, late fusion merges feature sequences from separate modules before the final stage, as evidenced in \cite{zhang2019feature, ji2020two, du2023frequency}. Hybrid fusion methods flexibly combine modality fusion and sequential modeling by evaluating inter-modality relationships, offering a versatile fusion strategy \cite{zhao2020trans2d, hu2023adaptive}.

\subsection{LLM for Recommendation}
The integration of Large Language Models into recommendation systems has been profoundly influenced by foundational models such as BERT \cite{devlin2018bert} and GPT-3 \cite{brown2020language}, which demonstrated the potential of LLMs in processing vast amounts of textual data to understand user behaviors deeply. This foundation has been expanded upon by subsequent models like BERT4Rec \cite{sun2019bert4rec} and innovations such as RLMRec \cite{ren2024representation}, which tailor LLM capabilities to generate personalized, context-aware recommendations by analyzing detailed user-item interactions.

In the current landscape, LLM applications in recommendation systems are categorized into three main approaches: embeddings-based, token-based, and direct model applications~\cite{wu2023survey, cao2024aligning, guo2024scaling}. Embeddings-based applications, such as \cite{cui2022m6,liu2024once}, use LLMs to extract rich feature representations from item and user data, enhancing the system's understanding of user preferences. Token-based approaches, highlighted in works like \cite{zhai2023knowledge}, focus on generating tokens that capture semantic meanings and potential user preferences, integrating this data into the recommendation logic. Lastly, direct model applications \cite{hou2024large, geng2022recommendation} involve using LLMs as end-to-end solutions where the models directly generate recommendations based on user queries and profiles, offering a streamlined and potentially more powerful system architecture. Additionally, MLLM-based recommendation frameworks have emerged to handle scenarios involving diverse data types like images, text, and video, improving system accuracy and user experience~\cite{liu2024rec, zhang2024notellm}.

%% file: sections/preliminary.tex
\section{Preliminary}
In this section, we will give the definition of our research problem and conduct a preliminary study to discuss the effectiveness of image summarizing approach.

\subsection{Problem Definition}
We first introduce the problem formulation of the Sequential Multimodal Recommendation problem. The dataset used in this work contains the interaction records between users and items. Given a user \(u\), let us first define the historical user behavior sequence of \(u\) as \(S_u = [I^1_u, \ldots, I^n_u]\), where \(I^i\) represents the \(i\)-th item with which the user has interacted, through actions such as clicking, purchasing, or watching, and \(n\) denotes the length of the user behavior sequence. In addition, each item corresponds to a textual description $\mathcal{W}$ and a image $\mathcal{I}$ (e.g., product diagram, video cover). Consequently, our problem can be formulated as follows.

\begin{definition}[Multimodal Sequential Recommendation]
    Given a user $u$ with the corresponding historical behavior sequence $S^u$, including both textual and visual data, and a candidate item $I_c$, the objective of the Multimodal Sequential Recommendation is to predict the probability of next interacted item $I^{n+1}_u$ (for example, the probability of clicking) with the candidate item $I^c_u$ for the user $u$, denoted as $g_u: I_c \rightarrow \mathbb{R}$.
\end{definition}

\subsection{Effectiveness of Multiple Images Summary}
As highlighted in the Introduction, current multimodal large language models (MLLMs) face challenges in processing multiple image inputs, limiting their effectiveness for sequential multimodal analysis. To overcome this problem, we introduce an image summary approach that leverages MLLMs to convert and summarize image content. The efficacy of this technique is evaluated using the basic sequential recommender, GRU4Rec, on real-world datasets (detailed in the Experiment section). In our approach, we employed simple prompts like "Please summarize the image" with LLaVA~\cite{liu2023llava, liu2024llavanext} to generate image summaries. These summaries were transformed into latent vectors using BERT~\cite{devlin2018bert}, which then fed into the GRU4Rec model. This method is benchmarked against direct image representations from VGG19~\cite{simonyan2014very}, assessing performance via the AUC metric.

\begin{table}[ht]
\centering
\caption{Performance of GRU4Rec with Different Inputs}
\begin{tabular}{cccc}
\hline
Dataset        & Microlens & Baby   & Games  \\ \hline

Image Summary  & 0.7281    & 0.7318 & 0.7451 \\ 

VGG19 Features & 0.7154    & 0.7383 & 0.7532 \\ \hline

\end{tabular}
\label{table:pre}
\end{table}
\vspace{2mm}

The performance is detailed in Table~\ref{table:pre}. Results show that using image summaries allows the GRU4Rec model to perform comparably to direct processing with VGG19, confirming that our image summary approach preserves necessary semantic information in sequential modeling. This preliminary validation underscores the effectiveness of our method in addressing the challenges associated with processing multiple ordered images.

%% file: sections/framework.tex
\begin{figure*}[t]
    \centering
    \includegraphics[width=0.85\linewidth]{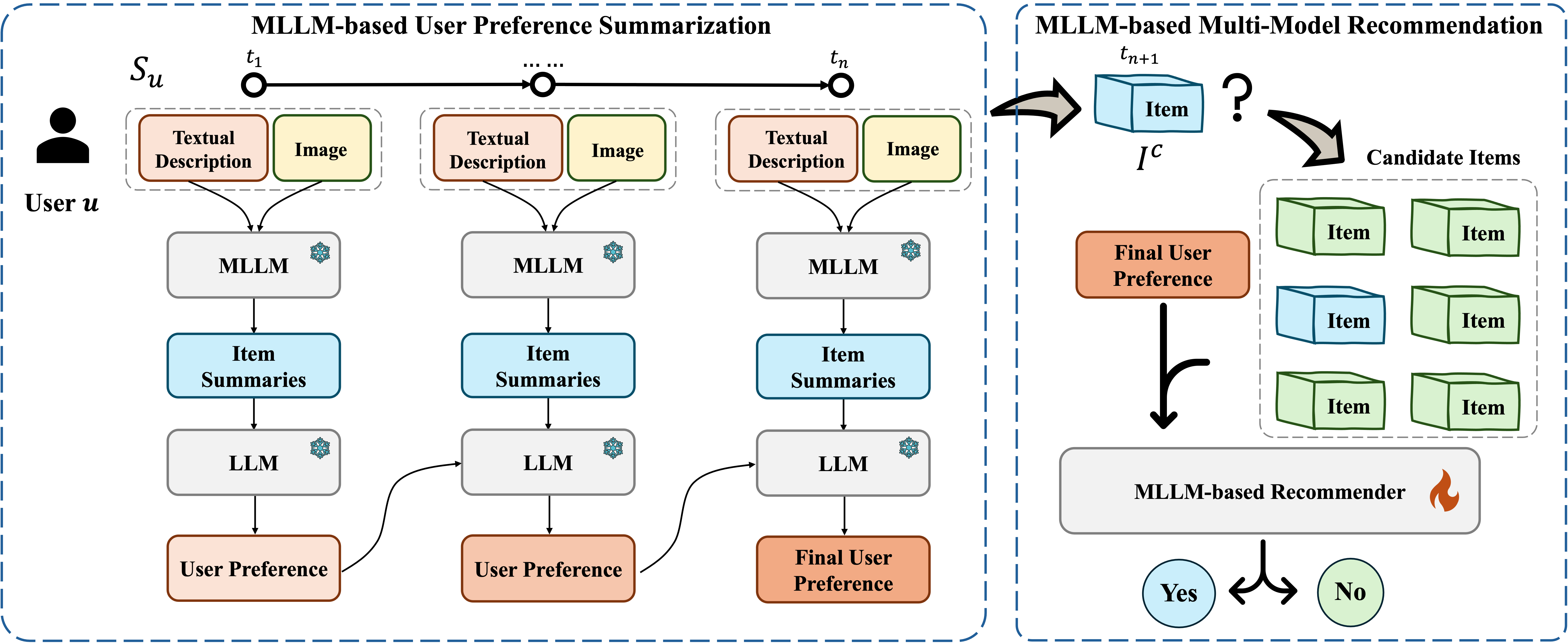}
    \caption{The schematic framework of MLLM-MSR}
    \label{fig:framework}
\end{figure*}

\section{Technical Details}
This section will introduce the technical details of our proposed MLLM-MSR framework, which contains two main components as Multimodal User Preferences Inference and Tuning MLLM based Recommender, illustrated as Figure~\ref{fig:framework}.

\begin{figure}[!hb]
    \centering
    \includegraphics[width=0.95\linewidth]{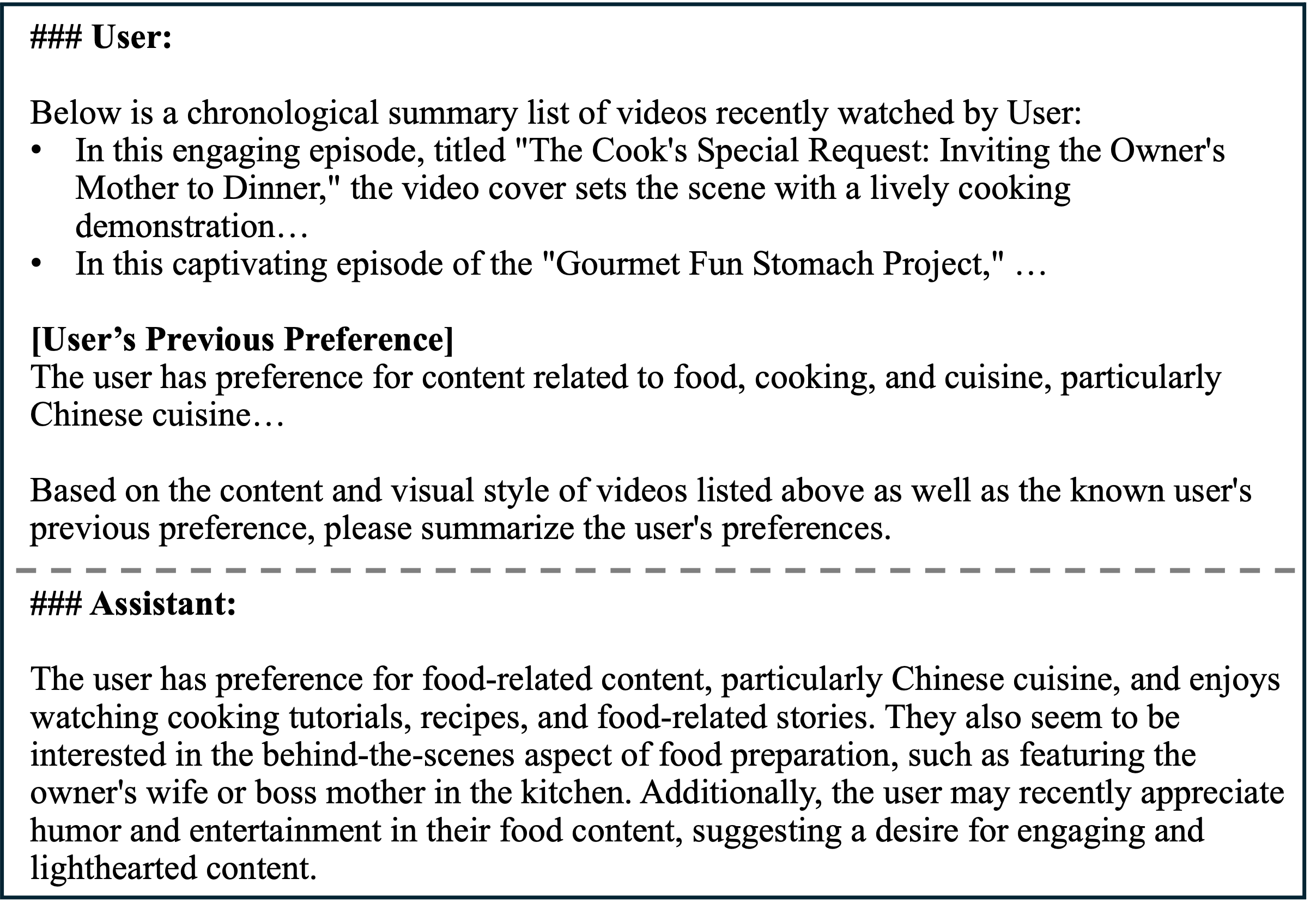}
    \caption{An example of recurrent user preference inference.}
    \label{fig:pref}
\end{figure}

\begin{figure*}[!ht]
    \centering
    \includegraphics[width=0.95\linewidth]{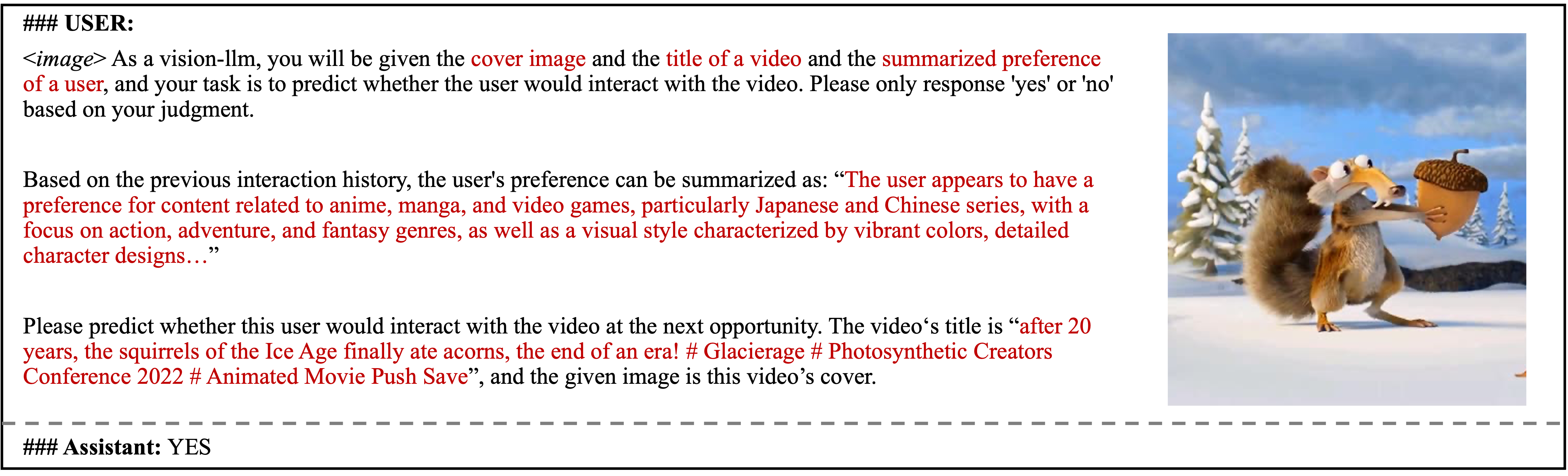}
    \caption{An example of MLLM-based sequential recommendation}
    \label{fig:example_rec}
\end{figure*}

\subsection{Multimodal User Preferences Inference}
In the context of sequential recommendation, a common approach is to learn user representations and predict future interactions with candidate items via calculating affinity scores. Unlike traditional methods that utilize embeddings, LLMs typically analyze user preferences and interaction probabilities directly at token level. This section will detail how our method employs Multimodal Large Language Models (MLLMs) to specifically address challenges associated with multimodal recommendation scenarios.

\subsubsection{Multimodal Item Summarization}
To effectively predict user preferences, it is crucial to analyze historical item sequences. In multimodal recommendation scenarios, handling multiple image inputs presents a significant challenge for MLLMs, especially in maintaining the sequence of these inputs and aligning textual information with corresponding images. To overcome these issues, we propose a Multimodal Item Summarization approach, which simplifies the processing by summarizing multimodal information of images into unified textual descriptions by designing effective prompts to integrate the multimodal data of items.

Our prompt design adheres to foundational methods of multimodal information fusion. Item information can be separated into textual descriptions and image. Hence, In the initial phase, distinct prompts (i.e., text summarization and image description prompt) are used to guide MLLMs to process these modalities independently, to ensure a more thorough comprehension and detailed feature extraction from each modality, ensuring nuanced characteristics often missed in unified analyses are captured. To ensure both modalities contribute equally to item modeling, the outputs of text summarization and image description are calibrated to similar lengths.

After independently analyzing each modality, our design integrates insights from both textual and visual information using a fusion prompt. This approach aligns with traditional multimodal recommendation strategies that emphasize synthesizing diverse data types to create a comprehensive item profile, enhances the multifaceted understanding of the item \cite{baltruvsaitis2018multimodal, huang2019acmmm, gao2020survey}.

\subsubsection{Recurrent User Preference Inference}
In the Sequential Multimodal Recommendation framework, achieving detailed personalization relies on an accurate understanding of user preferences. The advent of Multimodal Large Language Models (MLLMs) marks a significant advancement in understanding multimodal information. However, as we introduced above, they are struggle in dealing with sequential multimodal data. Although our multimodal item summarization method effectively integrates multimodal information into a unified item summary, this complexity still leads to unstable and random outputs when the historical sequence becomes long, leading to excessively long prompts. Consequently, this results in suboptimal performance in sequential recommendation systems.

To address these challenges, our method, inspired by Recurrent Neural Networks (RNNs), employs prompted sequence modeling to iteratively capture user preferences through interaction sequences. In RNNs, each output is influenced by both the current input and the previous state, facilitating contextual awareness across sequences. We segment item interactions into several blocks, each covering interactions within a defined session, converting long multimodal sequences into concise textual narratives that sequentially represent the user's historical interactions. This segmentation enables our approach to dynamically represent user preferences, effectively overcoming the limitations of MLLMs in processing sequential and multimodal data. By incorporating prompt-driven modules in each session, our method integrates insights from previous interactions to refine the understanding of current user preferences continuously. This iterative process is essential for accurately capturing the dynamic of user preferences and offers more interpretable descriptions than traditional representation-based models, enhancing the potential for detailed case studies.

Specifically, as we have initially generated item multimodal summaries for each item, we pair these summaries with prompts that guide the LLMs to infer user preferences based on a sequential narrative. For example, the initial prompt in the first block is designed to summarize the user's initial interests from a chronological list of item interactions at the first timestamp. Subsequently, at following session, the prompt for updating the summarized preference is showed as Figure~\ref{fig:pref}. Our approach uses prompted sequence modeling to iteratively understand user preferences through detailed analysis of each interaction, thus effectively manage challenges with long, multimodal sequences.

\subsection{Tuning MLLM based Recommender}
After gathering user preferences using the methods described above, we can propose a supervised fine-tuning of an open-sourced MLLM, such as LLaVA\footnote{https://huggingface.co/llava-hf/llava-v1.6-mistral-7b-hf}, which excels at understanding images. This model would be used to build a multimodal recommender system through SFT. In line with the definition of sequential recommendation, given a user-item interaction, the MLLM-based recommender system utilizes the prompt contained the obtained user preferences, the textual description and image of the given item and the designed system instruction prompt to predict the probability that the user will interact with the candidate item. Specifically, the prompt designed for the tuned MLLM recommender module is illustrated in Figure~\ref{fig:example_rec}, where we restrict the output to only include 'yes' or 'no' to avoid irrelevant information about the predicted label. Thus, the probability of item interaction can be calculated from the probability score of the predicted first new token as follows:

\begin{equation}
p = \frac{p(\text{'yes'})}{p(\text{'yes'}) + p(\text{'no'})}
\end{equation}

To construct a multimodal sequential recommender system based on MLLMs, we implement supervised fine-tuning to optimize the model parameters. This fine-tuning process involves adjusting the model to minimize the discrepancy between predicted and actual user interactions. Our dataset construction strategy employs negative sampling, a commonly used training technique in recommendation systems, wherein each positive user-item interaction is coupled with multiple negative samples representing items with which the user did not interact~\cite{yu2018multiple}. This methodology aids the model in distinguishing between relevant and irrelevant items through contrastive learning, thereby improving its predictive accuracy.

The model is trained on a dataset comprising sequences of user-item interactions, with each interaction encapsulated as a sequence of user preferences, item descriptions, and images. The fine-tuning leverages the next token prediction paradigm, training the model to predict the subsequent token in a sequence based on preceding tokens. This ensures the generation of coherent and contextually pertinent outputs from the input sequences. The supervised fine-tuning loss function is defined as:

\begin{equation}
    L = -\sum_{i=1}^L \log P(v_i|v_{<i}, \mathcal{I}),
\end{equation}
where $v_i$ represents the $i$-th token of the prompt text, $L$ denotes the prompt length and $\mathcal{I}$ is the given image. The probability $P(v_i|v_{<i}, \mathcal{I})$ is calculated using MLLMs within the next token prediction framework, which maximizes the likelihood of the ground truth tokens given the prompt. This ensures the model learns to accurately predict the subsequent token based on the provided context, which is critical for generating precise and contextually aware recommendations. Specifically, we employ LoRA~\cite{hu2021lora} to adhere to a parameter-efficient fine-tuning framework (PEFT), which accelerates the training process.

%% file: sections/experiments.tex
\section{Experiments}
In this section, we detail the comprehensive experiment to validate the effectiveness of our proposed Multimodal Large Language Model for Sequential Multimodal Recommendation (MLLM-MSR).

\renewcommand{\arraystretch}{1.1}
\begin{table}[!ht]
\small
\centering
\caption{The Statistics of Datasets}
\begin{tabular}{llll}
\hline
Dataset       & Microlens & Amazon-Baby & Amazon-Game \\ \hline
\#User        & 25411     & 41081       & 38808       \\
\#Item        & 20276     & 14393       & 13379       \\
\#Interaction & 223263    & 400876      & 352136      \\
\#Avg Seqlen & 11.35   & 13.65      & 13.23      \\
Sparsity      & 99.96\%   & 99.93\%     & 99.93\%     \\ \hline
\end{tabular}
\label{dataset}
\end{table}

\begin{table*}[!ht]
\centering
\small
\caption{The performance of different methods.}
\begin{tabular}{c|ccc|ccc|ccc}
\hline
         & \multicolumn{3}{c|}{Microlens}                                      & \multicolumn{3}{c|}{Video-Games}                                    & \multicolumn{3}{c}{Amazon-Baby}                                    \\
         & AUC                  & HR@5                 & MRR@5                & AUC                  & HR@5                 & MRR@5                & AUC                  & HR@5                 & MRR@5                \\ \hline
GRU4Rec  &   72.55                   &       53.32               &    33.36                  &        73.45              &       59.32               &        34.56              &      73.68                &         58.21             &       35.81               \\
SASRec   &74.02                   &58.88                      &36.57                      &71.06                      &43.81                      &23.94                      & 80.50                     &70.09                      &52.34                      \\

MGAT     &   71.23                   &     42.37                 &      30.84                &      72.21                &         48.71             &         32.17             &         73.19             &         46.42             &           31.92           \\
MMGCN    &    73.35                  &      54.47                &       34.14               &      74.91                &        55.63              &     35.19                 &         74.88             &        57.13              &        36.13              \\

GRU4Rec\textsubscript{F}  &75.31                   &      60.28                &     33.25                 &     75.79                 &     62.38                 &        36.39             &      76.41                &      63.84                &         35.77            \\
SASRec\textsubscript{F}   &77.69                      &63.03                      &37.42                     &77.51                      &66.64                      &43.55                      &81.94                      &72.08                      &58.32    
\\

MMSR   &      78.87                &      69.85                &     55.21                 &      79.01                &     71.95                 &   56.52                   &         81.19             &     70.34                 &             56.38         \\
Trans2D  &70.23                      &51.91                      &31.21                      & 72.20                     &  56.70                    & 39.28                     &66.60                      &28.94                      &45.91                      \\
LLaVa    & 52.75  & 28.21 & 17.33 & 53.57 & 33.24& 18.73 & 55.86& 34.82& 20.96\\
TALLREC  &     81.25                 &      71.23                &      58.22                &      81.79                &       72.36               &        57.38              &       82.16               &       74.31               &         60.15             \\
\hline
MLLM-MSR &    \textbf{ 83.17 }                &        \textbf{78.23 }             &     \textbf{60.38}                 &        \textbf{84.39  }            &          \textbf{77.42}            &         \textbf{63.25}             &        \textbf{85.69}              &         \textbf{79.58 }            &       \textbf{ 63.77   }           \\ \hline
\end{tabular}
\label{performance}
\end{table*}

\subsection{Experimental Setup}
\subsubsection{Dataset Description}
Our experimental evaluation utilized three open-source, real-world datasets from diverse recommendation system domains. These datasets include the \textit{Microlens Dataset}~\cite{ni2023content}, featuring user-item interactions, video introductions, and video cover images; the \textit{Amazon-Baby Dataset}; and the \textit{Amazon-Game Dataset}~\cite{he2016ups, mcauley2015image}, all of them contain user-item interactions, product descriptions, and images. These selections enabled a thorough analysis across different recommendation systems. We preprocessed each dataset by removing infrequent users and items to ensure user history sequences met our minimum length criteria. Additionally, we implemented a 1:1 ratio for negative sampling during training and a 1:20 ratio for evaluation. Further details on these datasets are provided in Table~\ref{dataset}.

\subsubsection{Baseline Methods} To evaluate the effectiveness of our proposed LMM-MSR method, we selected some compared methods which can be categorized into following groups:

\begin{itemize}
    \item \textbf{Basic SR Models}: These models use item attributes including IDs and textual information. We selectively integrate the most effective information from these attributes to achieve optimal performance. \textit{GRU4Rec~\cite{hidasi2015session}}: Utilizes Gated Recurrent Units (GRU) to model the sequential dependencies between items. \textit{SASRec~\cite{kang2018self}}: Employs a self-attention mechanism to capture long-term dependencies. 
    \item \textbf{Multimodal recommendation model}: \textit{MMGCN~\cite{wei2019mmgcn}}: Integrates multimodal features into a graph-based framework using a message-passing scheme. \textit{MGAT~\cite{monti2017mgat}}: Employs a graph attention network to disentangle personal interests by modality.
    \item \textbf{Multimodal feature enhanced SR models}: \textit{GRU4Rec\textsubscript{F}}, \textit{SASRec\textsubscript{F}}: Adaptations of GRU4Rec and SASRec with multimodal feature enhancements. \textit{Trans2D~\cite{zhao2020trans2d}}: Utilizes holistic fusion to integrate features across different dimensions.
    \textit{MMSR~\cite{hu2023adaptive}}: Depolys a graph-based approach for adaptive fusion of multi-modal features, which dynamically adjusts the fusion order of modalities based on their sequential relationships.
    \item \textbf{LLM based SR models}: \textit{TALLREC~\cite{bao2023tallrec}}: Uses LLMs for sequence recommendation through SFT, exclusively processing textual inputs. \textit{LLaVA w/o SFT}: Utilizes LLaVA as the recommender without a specific fine-tuning for recommendation.
\end{itemize}

\subsubsection{Metrics}
To assess the performance of baseline methods and our proposed MLLM-MSR for multimodal sequential recommendations, we employed AUC, HR@5, and MRR@5 as evaluation metrics~\cite{yu2020collaborative}. To ensure a fair comparison, we standardized the size of the candidate item sets across all baseline methods and our approach.

\subsubsection{Implementation Details}
Our experiments were performed on a Linux server equipped with eight A800 80GB GPUs. We utilized Llava-v1.6-mistral-7b for image description and recommendation tasks, and Llama3-8b-instruct~\footnote{{https://huggingface.co/meta-llama/Meta-Llama-3-8B-Instruct}} for summarizing user preferences. For the Supervised Fine-Tuning (SFT) process, we employed the PyTorch Lightning library, using LoRA with a rank of 8. The optimization was handled by the AdamW optimizer with a learning rate of 2e-5 and a batch size of 1, setting gradient accumulation steps at 8 and epochs at 10. For distributed training, we implemented Deepspeed [28] with ZeRO stage 2. Additionally, we set the maximum token length for MLLMs at 512 and the number of items per block in recurrent preference inference at 3.

\subsection{Performance Analysis}
The performance of the compared methods and our MLLM-MSR is presented in Table~\ref{performance}, where all results were obtained using 5-fold cross-validation and various random seeds, and achieved a 95\% confidence level. It is evident that, in our evaluation, MLLM-MSR consistently outperforms all other metrics in terms of both classification and ranking, underscoring the personalization accuracy of our recommendation system. We also observed additional insights:
Firstly, compared to basic sequential recommendation (SR) models, our adaptations that incorporate multimodal inputs, particularly SASRec, show significantly better results. This emphasizes the critical role of multimodal integration within the sequential recommendation framework and confirms the efficacy of self-attention-based models in handling both multi-modality and sequential inputs.
Moreover, MMSR distinguishes itself from other non-LLM baselines, highlighting the importance of integrating multimodal fusion modules with sequential modeling components in SR tasks, thereby indirectly supporting our prompt design idea for user preference inference.
Conversely, purely multimodal recommendation models such as MMGCN and MGAT exhibit lower performance due to their lack of dedicated sequential modeling components. This indicates that for optimal effectiveness in SR, the integration of both multimodal and sequential processing capabilities is essential.
Lastly, within the realm of large language model (LLM)-based SR models, our approach significantly outperforms LLaVA without specific fine-tuning. This success validates the effectiveness of our strategically designed prompts for SR tasks. Additionally, our method outperforms TALLREC, demonstrating our success in integrating multimodal information and unlocking the potential of large multimodal models compared to other LLM-based approaches using only textual information. This comparative advantage underscores the integration of advanced MLLM training techniques and the strategic application of multimodal data processing in enhancing sequential recommendation systems.

\subsection{Ablation Study}
To evaluate the individual contributions of certain components within our MLLM-MSR framework, we developed several variants of MLLM-MSR, each described as follows:

\begin{itemize}
\item \textbf{MLLM-MSR\textsubscript{R}}: This variant employs direct user preference inference instead of the recurrent method. It uses the entire chronological order of historical interaction data to infer user preferences.
\item \textbf{MLLM-MSR\textsubscript{$\mathcal{I}$}}: In this version, we omit the image summarization component in item summarization. It relies solely on textual data for user preference inference while still incorporating image information in the recommender module, leveraging capabilities that are readily achievable with current MLLMs.
\end{itemize}

\begin{figure}
    \centering
    \includegraphics[width=0.8\linewidth]{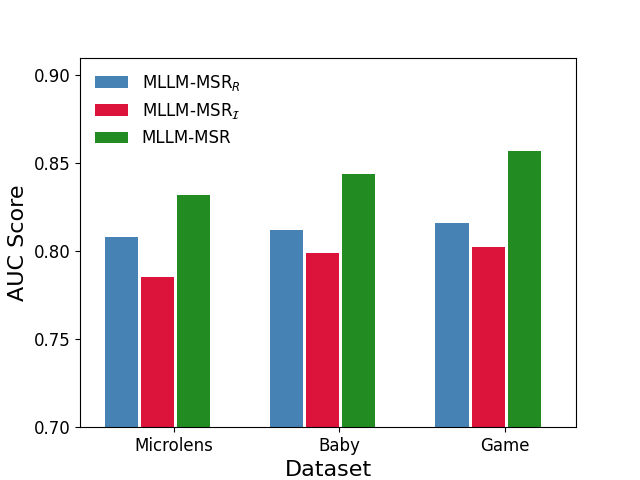}
    \caption{The performance of MLLM-MSR and its variants.}
    \label{fig:ablation}
\end{figure}

As Figure~\ref{fig:ablation} shows, in the evaluations across three diverse datasets, our primary model, MLLM-MSR, consistently outperformed its variants, MLLM-MSR\textsubscript{R} and MLLM-MSR\textsubscript{$\mathcal{I}$}, demonstrating the essential roles of its key components.
The MLLM-MSR\textsubscript{R} variant, which employed direct user preference inference, achieved suboptimal performance. This result validates the importance of our model's recurrent method in capturing the dynamic evolution of user preferences, indicates our methods can reflect current interests more accurately and reduce the negative impact of lengthy prompts. Besides, the worse performance of MLLM-MSR\textsubscript{$\mathcal{I}$} variant, which excluded image summaries and depended solely on textual data for user preference inference, illustrated the significance of integrating multimodal data. This integration is crucial to understand user preferences across different modalities, thereby significantly compensating for the incompleteness of textual information.

\begin{figure}[ht]
    \centering
    \begin{subfigure}{0.48\columnwidth}
    \includegraphics[width=\columnwidth]{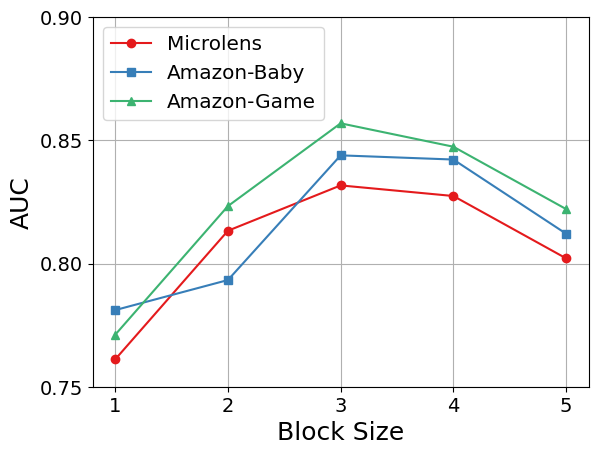}
    \end{subfigure}
    \begin{subfigure}{0.48\columnwidth}
        \includegraphics[width=\columnwidth]{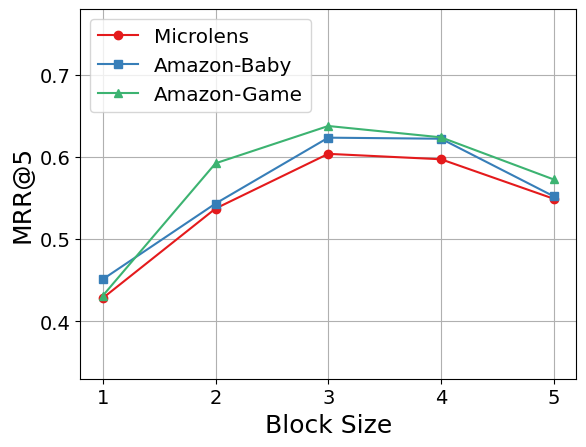}
		\end{subfigure}
	\caption{Performance of MLLM-MSR under different block size.}
	\label{fig:sp}
\end{figure}

\subsection{Parameter Analysis}
In this section, we first analyze the optimal block size for the recurrent user preference inference component of our MLLM-MSR model. As Figure~\ref{fig:sp} shows, striking the right balance on the block size is crucial; Too small a block size simplifies the approach to direct inference, potentially missing the dynamic evolution of user preferences due to the limited contextual span. Conversely, too large a block size leads to long prompts, increasing computational load and reducing the number of blocks available to effectively capture temporal dynamics, thereby diminishing the system's adaptive capabilities. Optimal block sizing ensures the model processes sequential data efficiently and adapts dynamically to changes in user behavior.

\begin{figure}[ht]
    \centering
    \begin{subfigure}{0.48\columnwidth}
    \includegraphics[width=\columnwidth]{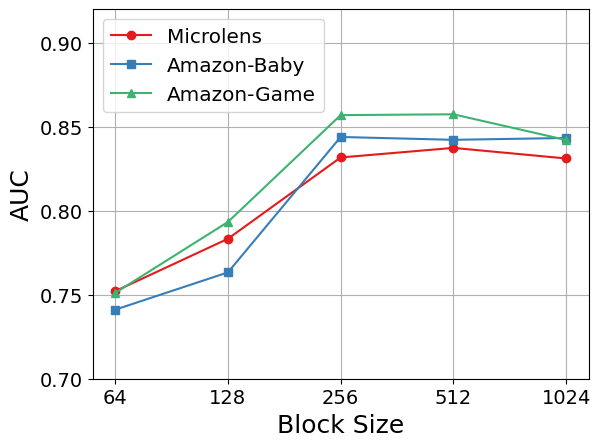}
    \end{subfigure}
    \begin{subfigure}{0.48\columnwidth}
        \includegraphics[width=\columnwidth]{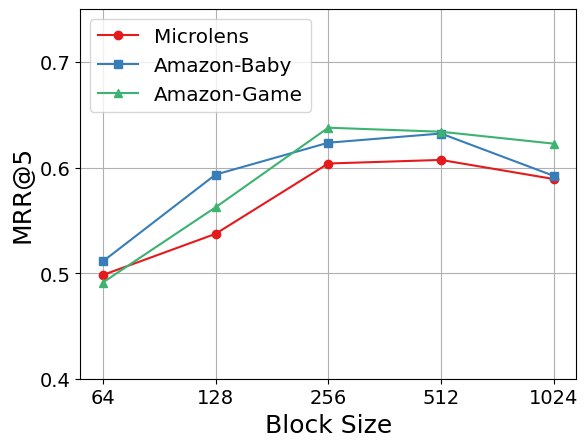}
		\end{subfigure}
	\caption{Performance of MLLM-MSR under different user preference summary length.}
	\label{fig:sl}
\end{figure}

Additionally, we evaluate the impact of context length on the predictive performance of our model.  By fixing the output length during user preference generation, we assess how different context lengths affect recommendation outcomes. The results are shown in Figure~\ref{fig:sl}. We found short context lengths cause a loss of information, resulting in suboptimal predictions. However, once the context length reaches a certain threshold, the results stabilize, indicating that the large model has strong summarizing capabilities and can capture all necessary information within a specific optimal range. This demonstrates the importance of selecting a proper context length to maximize information utility without incurring unnecessary computational complexity.

%% file: sections/conclusion.tex
\section{Conclusion}
In this study, our proposed model, the Multimodal Large Language Model-enhanced Multimodal Sequential Recommendation (MLLM-MSR), effectively leverages MLLMs for multimodal sequential recommendation. Through a novel two-stage user preference summarization process and the implementation of SFT techniques, MLLM-MSR showcases a robust ability to adapt to and predict dynamic user preferences across various datasets. Our experimental results validate the outstanding performance of MLLM-MSR compared to existing methods, particularly in its adaptability to evolving preferences. This paper introduces a innovative use of MLLMs that enriches the recommendation process by integrating diverse modalities and enhances the personalization and accuracy of the recommendations, and meanwhile providing added interpretability through detailed user preference analysis.